\documentstyle[mprocl]{article}

\begin{document}
\hfill TUW-00-25\\[1.5cm]

\title{2D QUANTUM GRAVITY WITH TORSION, DILATON THEORY AND 
BLACK HOLE FORMATION\footnote{Invited Talk in Parallel 
Session ``Metric Affine Gravity Models in 4 Dimensions'', 
9th Marcel Grossmann Conference, Rome, 2-8 July 2000}}

\author{W.\ KUMMER}

\address{Institut f\"ur Theoretische Physik, Technische 
Universit\"at Wien,\\ Wiedner Hauptstrasse 8-10/136, 
A-1040 Vienna, Austria\\
E-mail: wkummer@tph.tuwien.ac.at}


\maketitle
\abstracts{ 
Starting from the work of the author in 1990 with different
collaborators, essential progress in \( 2d \) gravity
theories has been made.  Now all such theories (and not
only certain special models) can be treated at the
classical as well as at the quantum level.  New physical
insights have been obtained, as e.g.  the {}``virtual black
hole{}''.  The formalism developed in this context recently
also finds increasing interest in mathematical physics. 
}

\section{Introduction}
The last years have seen an increased interest in dilaton
theories, treated in 1+1 dimensions.  The main motivation
to study such theories derives from the fact that spherical
reduction of D-dimensional Einstein gravity (SRG) precisely
generates a theory of this type \cite{Th}.

Consider a line element
\begin{equation}
\label{1}
\left( ds\right) ^{2}=g_{\mu \nu }dx^{\mu }dx^{\nu
}-\frac{2\left( D-2\right) \left( D-3\right) }{ 
\lambda^{2}}e^{-\frac{4}{D-2}\phi }\left( d\Omega_{D-2}\right)
^{2} 
\end{equation}
\noindent where \( d\Omega_{D-2} \) is the standard
surface element on the \( D-2 \) dimensional unit sphere \(
S^{D-2} \).  The dilaton field \( \phi \) depends on the
two first coordinates \( x^{\mu } \)only.  Then the \( D
\)-dimensional Einstein-Hilbert action, after \( \int \,
d\Omega _{D-2} \) has been dropped, becomes the one of SRG
\begin{equation}
\label{2}
{\cal L}_{SRG}=e^{-2\phi }\sqrt{-g}\left(
R+\frac{4\left( D-3\right) }{D-2}\left( \nabla \phi \right)
^{2}-\frac{\lambda ^{2}}{4}e^{\frac{4}{D-2}\phi }\right) ,
\end{equation}
\noindent (\( R \) is the \( 2d \) curvature scalar, \(
\nabla \phi ^{2}=g^{\mu \nu }\partial_{\mu }\phi 
\partial_{\nu }\phi \)) which represents a particular dilaton
theory with the Schwarzschild Black Hole (SBH) in \( D \)
dimensions as its general classical solution.

As of 1991 also the special case \( D\rightarrow \infty ,
\) (dilaton Black Hole , DBH) received particular attention
\cite{Ma}.  Inspired by string theory \cite{El} this model
also exhibits a BH solution, although its singularity is
null-complete \cite{Kat}.  On the other hand, (\ref{2})
allows a (classical) solution even when coupling to matter
is introduced.

The discussion of dilaton theories like (\ref{2}) also for
the more general case of \( \left( X/2=e^{-\phi }\right) \)
\begin{equation}
\label{3}
{\cal L}_{dil}=\sqrt{-g}\left( -\frac{X}{2}R+U\left(
X\right) \left( \nabla X\right) ^{2}-V\left( X\right)
\right) 
\end{equation}
\noindent for a long time had been the subject of a large
literature which, however, almost exclusively was
restricted to particular special cases \cite{dil}.  
Until the early 90-s (and sometimes even to this day) the
discussion of such dilaton models also was based
exclusively upon a particular gauge choice which indeed
plays a very prominent role in \( 2d \) gravity, the
conformal gauge \( \left( diag\, \eta =\left( 1_{\prime
}-1\right) \right) \)
\begin{equation}
\label{5}
g_{\mu \nu }=e^{2\rho }\, \eta_{\mu \nu }.
\end{equation}
\noindent However, although the resulting differential
equations for the dilaton field and \( \rho =\rho \left(
x\right) \) always turned out to be quite complicated, the
solution could be found in all cases.  As a relatively
simple example we may quote the Liouville equation for \(
\rho \) which appears in the JT model.  The choice of the
conformal gauge (\ref{5}) also was the reason why the DBH
could not be treated in full quantum theory - although
there with the action including a matter part
\begin{equation}
\label{6}
{\cal L}^{\left( m\right) }=\frac{1}{2}\sqrt{-g}\, F\left( X\right) 
\left( \nabla S\right) ^{2}
\end{equation}
\noindent for minimal coupling \( F\left( X\right) =1 \) to
the dilaton field, the exact solution was given in the
classical case.  However, the authors of \cite{El} and
others \cite{DBH2} only studied the semiclassical
approximation by including the one-loop scalars as an
effective Polyakov action \cite{Pol} in an otherwise
classical problem.

\section{Cartan variables and light cone gauge}

The history of the new approach started with the \( 2d \)
model of gravity with quadratic curvature and torsion which
was first formulated by Katanaev and Volovich \cite{KV} 
 (\( \varepsilon _{\mu \nu } \) and \(
\varepsilon_{ab} \) are the Levi-Civita symbol in
holonomic and anholonomic coordinates)
\begin{equation}
\label{7}
{\cal L}_{KV}=\sqrt{-g}\left( \frac{\alpha
}{2}R^{2}+\frac{\beta }{2}\tau_{a}\tau ^{a}-\Lambda
^{2}\right) 
\end{equation}
\begin{equation}
\label{8}
\tau ^{a}=\varepsilon ^{\mu \nu }\left( \partial_{\mu
}e^{a}_{\nu }+\omega_{\mu }{\varepsilon^a}_b\,
e^{b}_{\nu }\right) \end{equation}
\begin{equation}
\label{9}
-\sqrt{-g}\frac{R}{2}=\varepsilon ^{\mu \nu }\partial_{\mu
}\omega_{\nu }=r 
\end{equation}

As seen from (\ref{7}) - (\ref{9}) the presence of
nonvanishing torsion (\ref{8}) requires the formulation in
terms of Cartan variables \( e^{a}_{\mu } \) (zweibein) and
\( \omega ^{a}_{\mu }\, _{b}=\varepsilon ^{a}\, _{b}\,
\omega _{\mu } \) (spin connection, related to local
Lorentz transformations in the indices \( a=0,1 \)). 
Actually the consideration of non-vanishing torsion also in 
$d=4$ has a long history \cite{HE}. Motivated by the 
advantages of  ``physical'' (but noncovariant)
gauges like the temporal, axial or light-like gauges 
\cite{ax}, where at least the Faddeev-Popov complication is
absent, it seemed suggestive to use a light-like gauge for the
Cartan variables (\( e_{\mu }^{\pm
}=\frac{1}{\sqrt{2}}\left( e_{\mu }^{0}\pm e_{\mu
}^{1}\right) \))
\begin{equation}
\label{10}
e^{+}_{0}=e^{-}_{0}-1=\omega_{0}=0 \; .
\end{equation}
\noindent Then the \( 2d \) metric
\begin{equation}
\label{11}
g_{\mu \nu }=e^{a}_{\mu }\, e^{b}_{\nu }\, \eta _{ab} 
= e^{+}_{\mu }e^{-}_{\nu } + e_{\nu }^{+}\, e_{\mu }^{-}
\end{equation}
\noindent attains the Eddington-Finkelstein (EF) form
\begin{equation}
\label{12}
g_{\mu \nu }=e_{1}^{+}\left( 
\begin{array}{ll}
0 &\; 1\\
1 &\; e_1^-
\end{array}
\right)\; ,
\end{equation}
\noindent which - in contrast to the Schwarzschild type
diagonal or the conformal gauge - is well-known to be free
of coordinate singularities.  To the best of the author's 
this gauge had never been used before in the context of
quantum gravity - although its smoothness across the
horizon (zero of \( e_{1}^{-} \) for nonsingular \(
e^{+}_{1}>0 \)) is crucial for a dynamical treatment of a
full causally connected region in space-time.  In this
gauge all the quantum effects of the theory (\ref{7}) could
be lumped together in one divergent counter term.  The
finite quantum theory (after subtracting that term) just
coincided with the classical one \cite{KS1}.  This result
suggested to return to the classical theory and to repeat
the KV analysis in the gauge (\ref{10}).  Indeed the full
solution could be obtained after quite simple integrations 
\cite{KS2}.  Also the  path integral for this
model can be done exactly \cite{HK}: From the quadratic
dependence on the first time derivatives of fields in
(\ref{7}), both {}``momentum{}'' and
{}``coordinate{}''-integrals are of Gaussian type (as in
the linear oscillator) which permitted exact path
integrals. The Dirac quantization of that model was 
performed \cite{SS94}.

In a careful study of all generic \( 2d \) gravity theories
on the basic of the EF metric (\ref{12}) 
the possible global properties of
such theories were studied.  Again the absence of coordinate
singularities at the horizon(s) allowed to patch the
complete solutions in a \( C^{\infty } \) continuous way,
including also multiply connected spaces and solitonic
solutions \cite{KK96}.

\section{First order gravity}

Motivated by the first Hamiltonian analysis 
\cite{Grosse} and a first order formulation for the
(trivial) string case \cite{Verlinde} it was then realized 
\cite{TS94} that a first order \( 2d \) gravity action
\begin{eqnarray} L_{FOG} & = & \int\limits
_{{\cal M}}\left( X_{a}\, De^{a}+Xd\omega -\varepsilon
\, {\cal V}\right) =\nonumber \\ & = & \int d^{2}x\left(
X_{a}\tau ^{a}+Xr-\left( e\right) {\cal V}\right)
\label{13} 
\end{eqnarray}
\noindent with a general {}``potential{}'' \(
{\cal V}={\cal V}\left( X_{a}\, X^{a},X\right) \)
after elimination of \( X_{a} \) and \( X \) yields the
most general \( 2d \) gravity theory (with nonvanishing
torsion for \( \partial {\cal V}/\partial X^{2} \) \(
\neq 0 \)).  In the first line of Eq.  (\ref{13}) it is
written with differential one forms \( e^{a}=e_{\mu }^{a}\,
dx^{\mu } \) and \( \omega =\omega _{\mu \, }dx^{\mu } \)
for the Cartan variables.  The torsion two-form reads \(
De^{a}=de^{a}+\omega \varepsilon ^{a}_{b}\, e^{b} \) and
the curvature scalar is obtained from \( \left( e\right)
R=-2*d\omega =-2r \).  The volume two-form \( \varepsilon
=\frac{1}{2}e^{+}\wedge e^{-} \) is proportional to \(
\sqrt{-g}= \det\, e=\left( e\right) \).  The explicit
components in the second line are expressed in terms of \(
\tau _{a} \) and \( r \) as given in (\ref{8}) and
(\ref{9}).

Still, the connection with the physically motivated
(torsionless)dilaton theories (\ref{3}) or (\ref{2}) was
not established directly.  Of course, by a conformal
transformation \( \hat{g}_{\mu \nu }=g_{\mu \nu }exp\left(
-2\varphi \right) \) and thus
\begin{equation}
\label{14}
\sqrt{-\hat{g}\, }\hat{R}=\sqrt{-g}\, R
+ 2\, \partial_{\mu } 
\left( \sqrt{-g\, }g^{\mu \nu }\partial_{\nu }\varphi \right) 
\end{equation}
\noindent by a suitable choice for \( \varphi \), the
kinetic term \( \left( \nabla X\right) ^{2} \) in (\ref{3})
can be made to vanish \( \left( \hat{U}=0\right) \), and
the resulting model can be identified then directly with
(\ref{13}) for \( {\cal V}=\hat{V}\left( X\right) \) and
vanishing torsion.  However, that conformal transformation
not only locally (cf.  (\ref{14})), but also globally
completely changes the theory.  For example the DBH 
including minimal matter interactions in this way turns
into a theory of matter in a {\em flat} (Minkowski)
background.  All highly nontrivial dynamical interactions
between geometry and metric are eliminated.  Actually this
point has created much confusion in the literature because
in ordinary (quantum) gauge field theory a transformation
of the fields in the action (or a canonical transformation
in phase space) does not change the observables (S-matrix
elements).  In gravity the situation is completely
different, because the manifold upon which the matter
fields live does not remain unchanged (as it is the case
for Minkowski space in the flat theories) under a
transformation like the conformal one (\ref{14}), which is
not related to the symmetries of the theory
(diffeomorphisms, local Lorentz transformations).

The proof of local and global equivalence of a FOG theory
(\ref{13}) and a general dilaton theory (\ref{3}) emerged
in several steps.  It first was clarified for the KV model 
\cite{KKL}.  In its fully general form it is contained in
the work on \( 2d \) quantum gravity to be discussed below 
\cite{KLV}.  The argument is really quite straightforward. 
One first separates \( \omega =\widetilde{\omega }+\omega
^{\prime } \),   where \( \widetilde{\omega } \) is the
torsionless spin connection.  The quantity \( \omega
^{\prime } \) is proportional to \( X^{a} \).  Eliminating
\( X^{a} \) by its {\em algebraic} equations of motion
from the action (\ref{13}) directly yields the dilaton
theory (\ref{3}) when in (\ref{13}) the potential is
quadratic in \( X^{a}X_{a}=2X^{+}X^{-} \),
\begin{equation}
\label{15}
{\cal V}=\frac{X^{a}X_{a}}{2}U\left( X\right) +V\left( X\right) ,
\end{equation}
\noindent with the {\em same} functions \( U\left(
X\right) \) and \( V\left( X\right) \) as introduced in
(\ref{3}).  In that equation the (torsionless) \( R \)
simply depends on \( \widetilde{\omega } \).

At first glance the replacement of the torsionless dilaton
theory (\ref{3}) by an equivalent one with torsion
(\ref{13}) may seem an unnecessary complication, however it
is the first order form of (\ref{13}) which yields numerous
novel insights, especially for the quantum theory (to be
discussed below).  Already in the classical theory the
solution of the equations of motion for (\ref{13}) can be
given in a few lines for \( {\cal V} \) as in (\ref{15})
with arbitrary \( U \) and \( V \) (cf.\ \cite{KW1} for the
special case \( U=\alpha = \mbox{const} \))
\begin{eqnarray}
e^{+} & = & X^{+}e^{Q}df,\nonumber \\
e^{-} & = & \frac{dX}{X^{+}}+X^{-}e^{Q}df\; ,\label{16} \\
\omega  & = & -\frac{dX^{+}}{X^{+}}+Ve^{Q}df\; , \nonumber
\end{eqnarray}
\begin{eqnarray}
\label{17} 
C^{\left( g\right) } & = & X^{+}X^{-}e^{Q}+w\left( X\right) 
= \mbox{const.}\; ,\\
Q\left( X\right)  & = & \int ^{X}U\left( y\right) dy,\, w\left( X\right) = 
\int ^{X}e^{Q\left( y\right) }V\left( y\right) dy\; ,\nonumber 
\end{eqnarray}
\noindent where \( X \) and \( f \) are arbitrary
functions, closely related to the coordinates in the light
cone gauge of the last section.  Alternatively one arrives
at the solution also by the Darboux coordinates which can
be easily found here \cite{TS94}.  Eq.  (\ref{17})
represents an absolutely conserved quantity of the theory. 
It was known already (with \( Q=\alpha X \)) for the
matterless KV model, but, as shown first in \cite{KW1}, the
conservation law also holds when matter is present.  \(
{\cal C}^{\left( g\right) } \) generalizes the notion of
the ADM mass \cite{ADM} - valid for theories with
asymptotically flat solutions and systems with that
property - in a way which even allows a formulation
analogous to an {}``energy flux{}'' relation \cite{GK1}. 
It must be emphasized that the emergence of such a quantity
is an exclusive (but usually overlooked)
feature of \( 2d \) gravity theories and has important
consequences also e.g.  for the {}``physical{}'' SRG
(\ref{1}).  In the presence of matter the Noether symmetry
related to that conserved quantity turns out to be of a novel
type \cite{KT}.  The matter part possesses a conserved
one-form current \( J^{\left( m\right) } \) which provides
a new piece to be added to the (zero form) (\ref{17})
produced by the geometric part: The matterless conservation
law \( d\, {\cal C}^{\left( g\right) }=0 \) from the
equations of motion is found to be generalized to 
\cite{KW1}
\begin{equation}
\label{18}
d\; {\cal C}^{\left( g\right) }+J^{\left( m\right) }=0,
\end{equation}
\noindent where the matter current \( J^{\left( m\right) }
\) must obey \( d\, J^{\left( m\right) }=0, \) and thus
from Poincaré's Lemma \( J^{\left( m\right) }=d\,
{\cal C}^{\left( m\right) } \) and \( d\, \left(
{\cal C}^{\left( g\right) }+{\cal C}^{\left( m\right)
}\right) =0 \) follows.  The one-form conserved current \(
J^{\left( m\right) } \) introduces new symmetry parameters
which are independent from the ones in \(
{\cal C}^{\left( g\right) } \) \cite{KT}.

In connection with the material recounted so far in this
Section it must be emphasized that many of these, often
somewhat surprising new insights, are closely related to
the mathematic concept of Poisson Sigma Models (PSM) 
\cite{PSM}, which was a consequence of these developments. 
In the absence of matter an action like (\ref{13}) is just
a special case of more general class of \( 2d \) theory
defined by
\begin{equation}
\label{19}
L_{PSM}=\int\limits _{{\cal M}_{2}}\left[ dX^{I}\wedge A_{I} + 
\frac{1}{2}P^{IJ}A_{I}\wedge A_{J}\right]\; ,
\end{equation}
\noindent where \( X^{I} \) (in FOG: \( X^{I}=\left(
X,X^{a}\right) \)) are coordinates of a target space, and
\( A_{I} \) are \( 1 \)-form {}''gauge-fields{}'' (in FOG
:\( \, A_{I}=\left( \omega ,\, e_{a}\right) \)), both
dependent on the world sheet coordinates \( x^{0},\, x^{1}
\) of the manifold \( {\cal M}_{2} \).  The model is
defined by the singular Poisson structure \( P^{IJ}\,
\left( X\right) =-P^{JI}\, \left( X\right) \).  Whereas the
mixed components \( I=X,\, J=X^{a} \) determine the Lorentz
transformations, the component \(
P^{ab}={\cal V}\varepsilon ^{ab} \) is uniquely given by
that covariance and defines the specific model.  The
Poisson structure is related to a bracket operation \(
\left\{ X^{I},\, X^{J}\right\} =P^{IJ} \) of a (nonlinear)
algebra.  The ensuing Jacobi identity
\begin{equation}
\label{20}
P^{IL}\frac{\partial }{\partial X^{\L }}P^{JK} 
+ \mbox{cycl.\ perm.}\left( IJK\right) =0
\end{equation}
\noindent is the abstract reason for the symmetry
\begin{eqnarray}
\delta X^{I} & = & P^{IJ}\varepsilon_{J},\nonumber \\
\delta A_{I} & = & -d\, \varepsilon_{I} 
-\frac{\partial P^{JK}}{\partial X^{I}}\varepsilon 
_{K}A_{J}\; .
\label{21} 
\end{eqnarray}
\noindent PSM-s have found much interest not only in the
recent mathematical literature, but also for a formulation
of the star product suggested for string theory 
\cite{Schom}, where noncomutative geometry entered for the
first time.

Based on a proposal by Strobl \cite{TS99} in a (graded)
fermionic extension of (\ref{19}) a study of all possible
\( 2d \) supergravities (deformed by the dilaton field) is
in its final stages \cite{EKS}.

\section{\protect\( 2d\protect \) quantum gravity with matter}

The full impact of the advantages of the FOG form
(\ref{13}) of \( 2d \) gravity was realized in the
development of a quantum theory \cite{KLV} where, for the
first time, the geometric part could be integrated exactly
(in the path integral sense) for a arbitrary theory (\(
U\neq 0 \) in the potential (\ref{15})).  Matter can then
be treated in a systematic loop-wise expansion.  An
essential ingredient again is the light cone gauge
(\ref{10}).  It can be checked easily that (only) in that
gauge the Lagrangian in \( {\cal L}^{FOG} \) becomes
{\em linear} in the remaining geometric variables.

Of course, a careful study of the path integral has to
start from the Hamiltonian analysis, based upon the
existing powerful techniques of extended phase space
\cite{HK}, \cite{BV}.  It is convenient to introduce the
notation
\begin{eqnarray}
q_{i} & = & \left( \omega _{1},\, e_{1}^{+},\, e_{1}^{-}\right) ,\nonumber \\
\bar{q}_{i} & = & \left( \omega _{0},\, e_{0}^{+},\, e_{0}^{-}\right) 
\label{22} 
\end{eqnarray}
\noindent for the remaining dynamical variables of the
geometry (first line), when the variables in the second
line of (\ref{22}) have been fixed by (\ref{10}). 
Identifying \( x^{0} \) with {}``time{}'' in a natural
foliation of \( 2d \) space-time, the conjugate momenta to
\( q_{i} \) resp.  \( \bar{q}_{i} \) in (\ref{22}) are
\begin{eqnarray}
p_{i} & = & \left( X,\, X^{+},\, X^{-}\right) ,\nonumber \\
\bar{p}_{i} & \simeq  & 0.\label{23} 
\end{eqnarray}
\noindent The second line represents three primary
constraints.  The scalar field \( S \) from (\ref{6})
acquires a canonical momentum \( P=\partial
{\cal L}^{\left( m\right) }/\partial 
\left( \partial_{0}S.\right) \).

The canonical Hamiltonian follows as
\begin{equation}
\label{24}
H_{e}=\int dx^{1}\bar{q}_i\, G_{i},
\end{equation}
\noindent where the three (secondary {\em polynomial})
constraints \( G_{i}\left( p,q,P,S\right) \) not only in
the matterless \cite{Grosse}, \cite{HK} case, but also when
matter interactions by (\ref{6}) or with fermions 
\cite{K92} are present, fulfill the Poisson
brackets
\begin{equation} \label{25} \left\{ G_{i},\,
G^{\prime }_{j}\right\} ={\cal C}^{k}_{ij}\, G_{k}\,
\delta \left( x^{1}-x^{1\prime }\right) .  
\end{equation}
The structure functions \( {\cal C}^{k}_{ij}\, \left(
p\right) \) only depend on the momenta.  In contrast to the
well-known gravity algebras which are based upon the metric
(and not upon the Cartan variables as here) no derivative
acts on the \( \delta \)-function, i.e.\  (\ref{25}) is not
of Virasoro-type.  This greatly facilitates the extension
of the phase space (by two types of ghosts) \cite{BV}, the
construction of the BRS charge and of the gauge fermion,
appropriate for the gauge (\ref{10}).  The resulting path
integral in phase space has the structure
\begin{equation}
\label{26}
Z\left( j,J,Q\right) =\int \left( d\mu \right) 
\exp\, i\left( L_{\left( 1\right) }+L_{s}\right)\, ,
\end{equation}
\noindent where the measure
\begin{equation}
\label{27}
\left( d\mu \right) =\left( dp\right) \left(
d\bar{p}\right) \left( dq\right) \left(
d\bar{q}\right) \left( dP\right) \left( dS\right)
\left( d\left( \mbox{ghost} \right) \right) \sigma 
\end{equation}
\noindent also contains a suitable factor \( \sigma \)
which provides a correct final covariant integral for
matter (cf.\  the third ref.\ \cite{KLV}).  In the source
part of the action
\begin{equation}
\label{28}
L_{s}=\int \left( q_{i}\, j_{i}+p_{i}\, J_{i}+SQ\right) 
\end{equation}
\noindent beside sources \( j_{i}=\left(
j,j^{-},j^{+}\right) \) for \( q_{i} \) and \( Q \) for the
scalar field, also for convenience sources \( J_{i} \) for
the momenta are included.

After performing the (easy) integrals \( \left( d\left(
ghosts\right) \right) \left( d \bar{p}\right) \left(
d\bar{q}\right) \left( dP\right) \) Eq.(\ref{26})
turns into
\begin{equation}
\label{29}
Z = \int \left( d\widetilde{\mu }\right) 
\exp\, i\, \left( L_{\left( 2\right) }+L_{s}\right) 
\end{equation}
\noindent with the remaining integrals
\begin{equation}
\label{30}
\left( d\widetilde{\mu }\right) =\left( dp\right) 
\left( dq\right) \sqrt{\det\, e}\, \det\, F\left( dS\right) 
\end{equation}
\noindent and the factor det \( F \)
\begin{equation}
\label{31}
F=\partial _{0}+p_{2}U\left( p_{1}\right) 
\end{equation}
\noindent in the measure.  The (functional) determinant of
\( \left( e\right) =e_{1}^{+}=q_{3} \), left over from the
factor \( \sigma \) in (\ref{27}), ensures the covariance
of the final \( S \)-integral.  That term is lifted also
into the exponential by the identity
\begin{equation}
\label{32}
\sqrt{\det\, q_{3}}=\int \left( d\varphi \right) \left( dc\right) 
\left( dc^{\prime }\right) \exp\, i\, \int q_{3}\left( \varphi^{2} 
+ c^{\prime }c\right)\; ,
\end{equation}
\noindent where \( \varphi \) is a commuting, \( c \) and
\( c^{\prime } \) are anticommuting auxiliary field
variables.  Now - as announced at the beginning of this
Section - the remaining effective action \( L_{\left(
2\right) } \) together with the factor (\ref{32})
{\em only contains terms independent of or linear in \(
q_{i} \)}, appearing in the combinations
\begin{eqnarray}
{\cal L}_{\left( 2\right) } & = & q_{i}\left( \partial
_{0}p_{1}+A_{1}\right) +q_{2}\left( \partial
_{0}p_{2}+A_{2}\right) +\nonumber \\ & & +q_{3}\left(
Fp_{3}+A_{3}\right) +(terms\, without\, q_{i})\; .
\label{33}
\end{eqnarray}
\noindent The quantities \( A_{i} \) depend on \( j_{i,}\,
p_{i} \) and \( (\partial_{o}S)^{2} \).  We restrict
ourselves in the following to the simpler case of minimally
coupled scalars, where (\ref{6}) may be rewritten as
\begin{equation}
\label{34}
{\cal L}^{\left( m\right)
}=-\frac{1}{2}\frac{\varepsilon ^{\alpha \mu }\,
\varepsilon ^{\beta \nu }}{\left( e\right) }\, e_{\mu
}^{a}e_{\nu \, }^{b}\eta_{ab}\left( \partial_{\alpha
}S\right) \left( \partial_{\beta }S\right) ,
\end{equation}
\noindent which also in the gauge (\ref{10}) only yields
linear terms in \( q_{i} \).  Contrary to the usual
sequence of phase space integrals, now {\em first} \( \int
\left( dq\right) \) is performed.  The resulting \( \delta
\)-functions from (\ref{33}) determine \( p_{i} \) as the
solution of the (classical!) e.o.m.-s for \( p_{i} \).  The
solutions of these linear differential equations contain
inhomogeneous parts \( \bar{p}_i \) with \( \partial_0 
\bar{p}_i=0 \) which are found to yield
precisely the (classical) geometric background.  For
example in SRG the Schwarzschild black hole appears if \(
\bar{p}_3\neq 0 \).  From the integration \( \left(
d^{3}p\right) \) those solutions \( p_{i}=B_{i} \) are
simple reinserted everywhere, and the trick (\ref{32}) is
performed backwards:
\begin{equation}
\label{35}
Z=\int \left( dS\right) \, \, \sqrt{\det\, E^{+}_{1}}\,
\exp\, i\, L_{\left( 3\right) } 
\end{equation}
\begin{equation}
\label{36}
L_{\left( 3\right) }=\int \left( J_{i}\, B_{i}+SQ+\left(
\partial _{0}S\right) \left( \partial _{1}S\right)
+\widehat{{\cal L}}\right) 
\end{equation}

>From (\ref{33}) and (\ref{30}) the determinant $F$ cancels 
so that -- as expected -- no Faddeev-Popov determinant 
survives. 
In (\ref{35}) \( E^{+}_{1} \) becomes nothing else than the
classical solution for \( e^{+}_{1} \), expressed in terms
of \( B_{i} \) (which beside the gravity sources \( j_{i}
\) contains the scalar field).  Thus the full back reaction
is taken into account.  In the effective action the
insertion of \( B_{i} \) for \( p_{i} \) in the source term
(\ref{28}) does not only produce the first term in
(\ref{36}).  Since \( B_{i} \) contains terms like \(
\partial ^{-1}_{0}G \) which, by partial integration from a
term \( \partial_{0}\bar{g}=0 \) added to \( J_{i}
\), may yield a nontrivial contribution prop.  \(
\bar{g} G \).  This term thus survives the limit \(
J_{i}=0 \), i.e.\  even becomes the most important one for a
{}``physical{}'' Green-function without external momentum
lines.  It was first discussed for the KV model in 
\cite{HK}.  As shown in \cite{KLV} it is responsible for an
exact reproduction of the classical solution to the
zweibeine \( q_{i} \) when matter is switched off, here, of
course, it contains the full dependence on the matter
fields through a dependence on \( \left( \partial_{0} S
\right)^{2}.  \)

The last path integral cannot be done exactly any more. 
However, a loop-wise expansion in \( S \) is derived from
expanding \( \widehat{{\cal L}} \) and \( B_{i} \) in
powers of \( \left( \partial_{0}S\right)^{2} \).  For the
quadratic part \( O\left( S^{2}\right) \) in (\ref{36}) the
path integral is Gaussian.  It may be expressed as a
Polyakov type action plus a propagator term \( i\int_{x}\,
\int _{y}\, Q\left( x\right) \Delta ^{\left( SS\right)
}_{\, xy}Q \) \( \left( y\right) /2 \) in the exponent. 
The higher (even) powers \( O\left( S\right) ^{2n} \) \(
\left( n>2\right) \) yield effective vertices for the loop
expansion.  As sketched in the last ref.\ \cite{KLV} and
shown recently \cite{GKV} the vertex \( O\left( 
S\right)^{4} \) -- without any loop corrections! -- in an \( S
\)-matrix element of asymptotically flat space for the
scattering of two scalars \( S+S\rightarrow S+S \) for SRG
proceeds through an intermediate {}``virtual black
hole{}''.
\vspace*{-2pt}

\section{Conclusions and outlook}

Quantum gauge theories in two dimensions have become a
mature field in the last decade.  Contrary to the belief 
ten years ago, in the
absence of matter {\em all} \( 2d \) gravities are now
well understood as {}``almost{}'' topological theories. 
With matter interactions a highly nontrivial quantum
gravity emerged which allowed the treatment of spherically
reduced gravity with its direct relation to \( 4d \)
Einstein gravity.  The future plans include the 
investigation of 
higher loops, of the dynamics of supersymmetric extensions
and, as a new very promising field, the application to
spherically reduced generalized \( 4d \) Einstein theories,
which as {}``quintessence{}'' theories have brought about a
revival of the Jordan-Brans-Dicke theories \cite{GHK}.

\section*{Acknowledgements}

The author is grateful to his collaborators in the quoted 
literature. This research was supported by several projects 
of the Fonds zur 
F\"orderung der Wissen\-schaft\-lichen Forschung (Austrian 
Science Foundation), the most recent one being 
P12815-TPH.


\vspace*{-9pt}

\section*{References}

\begin{thebibliography}{10}

\bibitem{Th}P.\  Thomi, B.\  Isaak and P.\  Hajicek,
Phys.Rev.\  D {\bf 30} (1984) 1168; 
P.\  Hajicek, Phys.\ Rev.\  D {\bf 30}
(1984) 1178. 

\bibitem{Ma} G.\  Mandal, A.\  Sengupta and S.R.\  Wadia, Mod.\ 
Phys.\  Lett A {\bf 6} (1991) 1685; V.P.\  Frolov,
Phys.\  Rev.\  D {\bf 46} (1992) 5383; J.G.\  Russo
and A.A.\  Tseytlin, Nucl.\  Phys.\  B {\bf 382}
(1992) 259; J.G.\  Russo, L.\  Susskind and L.\  Thorlacius,
Phys.\  Lett.\  B {\bf 292} (1992) 13; T.\  Banks,
A.\  Dabholkar, M.\  Douglas and M.\  O'Loughlin, Phys.\  Rev.\ 
D {\bf 45} (1992) 3607; S.P.\  deAlwis, Phys.\ 
Lett.\  B {\bf 289} (1992) 278; S.D.\  Odintsov and
I.L.\  Shapiro, Phys.\  Lett B {\bf 263} (1991)
183; E.\  Elizalde, P.\  Fosalba-Vela, S.\  Naftulin and S.D.\ 
Odintsov, Phys.\  Lett.\  B {\bf 352} (1995) 235.

\bibitem{El}S.\  Elitzur, A.\  Forge and E.\  Rabinovici,
Nucl.\  Phys.\  B {\bf 359} (1991) 581; E.\  Witten,
Phs.\  Rev.\  D {\bf 44} (1991) 314; C.G.\  Callan,
S.B.\  Giddings, J.A.\  Harvey and A.\  Strominger, Phys.\ 
Rev.\  D {\bf 45} (1992) 1005.

\bibitem{Kat}M.O.\  Katanaev, W.\  Kummer and H.\  Liebl,
Phys.\  Rev.\  D {\bf 53} (1996) 5609; M.O.\ 
Katanaev, W.\  Kummer and H.\  Liebl, Nucl.\  Phys.\  B
{\bf 486} (1997) 353.

\bibitem{dil}Without completeness we just cite: 
C.\  Teitelboim, Phys.\  Lett.\ 
{\bf 126} B (1983) 41; R.\  Jackiw,
{\it Quantum Theory of Gravity}, ed.\  S.\  Christensen p.\ 
403,  Hilger, Bristol 1984; 
E.\ d'Hoker, D.Z.\  Freedman and R.\  Jackiw, Phys.\  Rev.\  D.\ 
{\bf 28} (1983) 2583; R.\  Jackiw, Nucl.\  Phys.\  B
{\bf 252} (1985) 343; D.\  Cangemi and R.\  Jackiw,
Phys.\  Rev.\  Lett.\  {\bf 69} (1991) 233; R.B.\ 
Mann, A.\  Shiekh and L.\  Tarasov, Nucl.\  Phys.\  B
{\bf 341} (1990) 134; D.\  Banks and M.\ O'Loughlin,
ibid.\  B {\bf 362} (1991) 649; H.J.\  Schmidt, J.\ 
Math.\  Phys.\  {\bf 32} (1991) 1562; V.P.\  Frolov,
Phys.\  Rev.\  D {\bf 46} (1992) 5383; J.G.\  Russo
and A.A.\  Tseytlin, Nucl.\  Phys.\  B {\bf 382}
(1992) 259; I.V.\  Volovich, Mod.\  Phys.\  Lett.\  A
{\bf 8} (1992) 1827; R.P.\  Mann, Phys.\  Rev.\  D
{\bf 47} (1993) 4438; D.\  Louis-Martinez, J.\ 
Gegenberg and G.\  Kunstatter, Phys.\  Lett.\  B
{\bf 321} (1994), 193; D.\  Louis-Martinez and G.\ 
Kunstatter, Phys.\  Rev.\  D {\bf 49} (1994) 5227.

\bibitem{DBH2}J.G.\  Russo, L.\  Susskind and L.\ 
Thorlacius, Phys.\  Rev.\  D {\bf 46} (1993) 3444, D
{\bf 47} (1993) 533.

\bibitem{Pol}A.M.\  Polyakov, Phys.\  Lett B
{\bf 103} (1981) 207.

\bibitem{KV}M.O.\  Katanaev and I.V.\  Volovich, Phys.\ 
Lett.\  B {\bf 175} (1986) 413; M.O.\  Katanaev and I.V.\ 
Volovich, Ann.\  Phys.\  {\bf 197} (1990) 1.

\bibitem{HE}F.W.\  Hehl in {\it Proc.\  6th School of
Cosmology and Gravitation on Spin, Torsion, Rotation and
Supergravity,} (P.G.\  Bergmann and V.\  deSabbata eds.) Nato
Advanced Study Institute, Erice 1978, Vol.\ 
{\bf 58} of Series B: Physics, 1980 .

\bibitem{ax}W.\  Kummer, Acta.\  Phys.\  Austriaca
{\bf 14} (1961) 149; {\bf 41}
(1975) 315; {\bf 52} (1980) 141; W.\  Konetschny
and W.\  Kummer, Nucl.\  Phys.\  B {\bf 100} (1975)
106, B {\bf 108} (1976) 397, B
{\bf 124} (1977) 145; W.\  Kummer and M.\  Schweda,
Phy.\  Lett.\  {\bf 141} B (1984) 363.

\bibitem{KS1}W.\  Kummer and D.J.\  Schwarz, Nucl.\  Phys.\ 
B {\bf 382} (1992) 171.

\bibitem{KS2}W.\  Kummer and D.J.\  Schwarz, Phys.\  Rev.\ 
D {\bf 45} (1992) 3628.

\bibitem{HK}F.\  Haider and W.\  Kummer, Int.\  Journ.\ 
Mod.\  Phys.\  A {\bf 9} (1994) 207.

\bibitem{SS94}P.\  Schaller and T.\  Strobl, Class.\ 
Quant.\  Grav.\  {\bf 11} (1994) 331.


\bibitem{KK96}T.\  Kl\"osch and T.\  Strobl, Class.\  Quant.\ 
Grav.\  {\bf 13} (1996) 965, {\bf 14}
(1997) 1689.

\bibitem{Grosse}H.\  Grosse, W.\  Kummer, P.\  Presnajder,
D.J.\  Schwarz, Journ.\  Math.\  Phys.\  {\bf 33}
(1992) 3892.

\bibitem{Verlinde}H.\  Verlinde, in {\it 6th Marcel
Grossmann Meeting on General Relativity}, Kyoto, Japan
(1991) (M.\  Sato ed.) World Scientific 1992, p.\  830; N.\ 
Ikeda and K.-I.\  Izawa, Progr.\  Theor.\  Phys.\ 
{\bf 89} (1993) 223.

\bibitem{TS94}T.\  Strobl, Phys.\  Rev.\  D {\bf 50}
(1994) 7356.

\bibitem{KKL}M.O.\  Katanaev and W.\  Kummer, H.\  Liebl,
Phys.\  Rev.\  D {\bf 53} (1996) 5609, Nucl.\  Phys.\  B
{\bf 486} (1997) 353.

\bibitem{KLV}W.\  Kummer, H.\  Liebl and D.V.\ 
Vassilevich, Nucl.\  Phys.\  B {\bf 493} (1997) 491, B
{\bf 513} (1998) 723, B {\bf 544} (1999) 403.

\bibitem{KW1}W.\  Kummer and P.\  Widerin, Phys.\  Rev.\  D
{\bf 52} (1995) 6965.

\bibitem{ADM}R.\  Arnowitt, S.\  Deser and C.\ W.\  Misner,
in {\it Gravitation: An Introduction to Current
Research} (L.\  Witten, ed.), Wiley, New York 1962.

\bibitem{GK1}D.\  Grumiller and W.\  Kummer, Phys.\  Rev.\ 
D {\bf 61} (2000) 064006.

\bibitem{KT}W.\  Kummer and G.\  Tieber, Phys.\  Rev.\  D
{\bf 59} (1999) 044001-1.

\bibitem{PSM}P.\  Schaller and T.\  Strobl, Mod.\  Phys.\ 
Lett.\  A {\bf 9} (1994) 3129.

\bibitem{Schom}V.\  Schomerus, JHEP {\bf 9906} (1999)
030; N.\  Seiberg and E.\  Witten, JHEP {\bf 9909} (1999)
032.

\bibitem{TS99}T.\  Strobl, Phys.\  Lett.\  B {\bf 460}
(1999) 87.

\bibitem{EKS}M.\  Ertl, W.\  Kummer and T.\  Strobl (in
preparation).

\bibitem{BV}E.S.\  Fradkin and G.A.\  Vilkovisky, Phys.\ 
Lett.\  B {\bf 55} (1975) 224; I.A.\  Batalin and G.A.\ 
Vilkovisky, Phys.\  Lett.\  B {\bf 69} (1977) 309; E.S.\ 
Fradkin and T.E.\  Fradkina, Phys.\  Lett.\  B {\bf 72}
(1978) 343; cf.\  also M.\  Henneaux, Phys.\  Rep.\ 
{\bf 126} (1985) 1; J.\  Govaerts {\it Hamiltonian
Quantization and Constrained Dynamics,} McGraw Hill, New
York 1965 p.\  1.

\bibitem{K92}W.\  Kummer in {\it Hadron Structure
'92} (D.\  Brunsko and J.\  Urban, eds.) Kosice 1992, p.48;
H.\  Pelzer and T.\  Strobl, Class.\  Quant.\  Grav.\ 
{\bf 15} (1998) 3803.

\bibitem{GKV}D.\  Grumiller, W.\  Kummer and D.V.\ 
Vassilevich, Nucl.\  Phys.\  B (to be published).

\bibitem{GHK}D.\  Grumiller, D.\  Hofmann and W.\  Kummer,
Ann.\  of Phys.\  (NY), (to be published).

\end{thebibliography}
\end{document}